\documentclass[amsmath,trackchanges,twocolumn]{aastex702}
\usepackage[T1]{fontenc}

\shorttitle{LSP Phase Offsets}
\shortauthors{Skowron et al.}

\submitjournal{ApJ}

\begin{document}

\title{Phase Offsets Between the Minimum Light and Radial Velocity in Long Secondary Period Stars}

\author[0000-0001-9439-604X]{Dorota M. Skowron}
\affiliation{Astronomical Observatory, University of Warsaw, Al. Ujazdowskie 4, 00-478 Warsaw, Poland}
\email[show]{dszczyg@astrouw.edu.pl}

\author[0000-0002-7777-0842]{Igor Soszy{\'n}ski}
\affiliation{Astronomical Observatory, University of Warsaw, Al. Ujazdowskie 4, 00-478 Warsaw, Poland}
\email{soszynsk@astrouw.edu.pl}

\author[0000-0002-6212-7221]{Patryk Iwanek}
\affiliation{Astronomical Observatory, University of Warsaw, Al. Ujazdowskie 4, 00-478 Warsaw, Poland}
\email{piwanek@astrouw.edu.pl}

\author[0000-0003-2244-1512]{Piotr A. Kołaczek-Szymański}
\affiliation{Astronomical Observatory, University of Warsaw, Al. Ujazdowskie 4, 00-478 Warsaw, Poland}
\affiliation{Astronomical Institute, University of Wrocław, Mikołaja Kopernika 11, 51-622 Wrocław, Poland}
\email{piotr.kolaczek-szymanski@uwr.edu.pl}

\begin{abstract}
Long secondary periods (LSPs), observed in a substantial fraction of luminous red giants, remain one of the longstanding problems in red giant variability. Previous studies have found a systematic phase relation between their photometric and
radial velocity (RV) variations. We investigate this relation using ASAS-SN photometry and {\it Gaia} RV curves for 983 LSP stars, and compare it with other types of Long Period Variables. We confirm that LSP stars have a relatively narrow distribution of phase offsets between the phase of minimum RV and minimum light, with a mean value of $\Delta\phi=-0.15\pm0.10$ in phase ($-0.96\pm0.64$ rad).
Interestingly, this phase offset is similar for Miras, showing that this quantity is not sufficient to distinguish LSPs from pulsational variability. 
However, pulsating red giants show a relation between period and RV amplitude, with the RV amplitude generally increasing toward longer periods. LSP stars deviate from this relation, showing RV amplitudes of only a few km~s$^{-1}$ despite their long periods.
The stacked light and RV curves of the LSP sample also show remarkable homogeneity and a coherent phase relation. In the binary interpretation of LSPs, the observed phase offset requires the obscuring material to be shifted in orbital phase with respect to the companion, consistent with models in which a low-mass companion produces a dusty wake in the wind of the red giant.
\end{abstract}


\section{Introduction}\label{sec:introduction}

Long secondary periods (LSPs) are a common but still unexplained form of variability in luminous red giants. First reported nearly a century ago by \cite{oconnell1933}, LSPs are photometric variations with periods of several hundred to several thousand days, typically about an order of magnitude longer than the primary pulsation periods of red giant branch (RGB) and asymptotic giant branch (AGB) stars. Their amplitudes can reach $\sim$1 mag in the optical bands, while associated radial-velocity (RV) variations have full amplitudes of a few~km~s$^{-1}$ \citep{hinkle2002,wood2004,nicholls2009}. In the period--luminosity diagram, LSP stars form the well-known sequence D \citep{wood1999,wood2000}. The phenomenon is remarkably common, occurring in approximately one third of pulsating red giants \citep{soszynski2021}, but the physical origin of LSPs remains uncertain.

In recent years, the debate over the origin of LSPs has gained renewed attention, with new large-scale observational studies and theoretical investigations providing increasingly stringent constraints on the mechanisms proposed to explain the phenomenon \citep{goldberg2024,pawlak2024,decin2025,macleod2025,trabucchi2025,courtneybarrer2026,iwanek2026,kolaczek2026,shariat2026}. Their strongly non-sinusoidal light curves are often accompanied by infrared excesses and evidence for asymmetric circumstellar dust \citep{wood2009,pawlak2021}. A variety of single-star mechanisms have been proposed, including non-radial oscillatory convective modes \citep{saio2015,takayama2020}, large-scale convective structures \citep{stothers2010}, mode interactions, rotational or magnetic modulation \citep{soker1999, wood2004}, and variability associated with episodic dust formation and mass loss \citep{wood2004,takayama2015,mcdonald2019}. However, no single-star mechanism has yet reproduced the full range of observed photometric, spectroscopic, and infrared properties.

An alternative interpretation associates the LSP with a low-mass companion interacting with the extended atmosphere and wind of the red giant. The companion may generate an asymmetric dusty cloud or trailing wake that periodically obscures the star \citep{wood1999,soszynski2014,soszynski2021}. The detection of secondary minima in the mid-infrared light curves of a substantial fraction of LSP stars provides additional support for this scenario \citep{soszynski2021}, while the observed RV amplitudes are broadly consistent with companions in the brown-dwarf or very-low-mass stellar regime \citep{nicholls2009}.

The RV variations, however, reveal properties that are not readily explained by a simple binary model. The non-sinusoidal shapes of RV curves suggest eccentric orbits, but Keplerian fits yield a pronounced concentration of the argument of periastron at $\omega>180^\circ$, with a median of approximately $227^\circ$, rather than the uniform distribution expected for randomly oriented orbits \citep{wood2004,nicholls2009}. A systematic phase relation between the RV and optical light curves has also been reported. \citet{nicholls2009} found that the phase of minimum light in their LSP sample occurs close to the phase of rising RV, while \citet{goldberg2024} found the corresponding RV--$I$-band phase difference to cluster around $\Delta\phi\sim-\pi/2$. The physical origin of these relations remains unclear.

\citet{decin2025} proposed that both effects may arise from the interaction of an eccentric, close-in low-mass companion with the dust-formation region of the giant. In their model, the companion spends only part of its orbit in the dust-forming wind and gravitationally focuses the outflow into an asymmetric wake. This geometry can produce a preferential distribution of $\omega$ and, according to their three-dimensional hydrodynamical simulations, shift the maximum optical depth of the wake downstream from the companion by approximately $90^\circ$--$225^\circ$ in orbital plane. The model therefore provides a possible explanation for the observed RV--light-curve phase relation as well as the $\omega>180^\circ$ excess \citep{decin2025}.

The phase relation reported by \citet{nicholls2009} and investigated further by \citet{goldberg2024} thus provides an important observational constraint on the binary--dust interpretation of LSPs. However, these studies were based on relatively small samples. In this work, we investigate the phase relation between the LSP light curves and RV variations for a substantially larger sample of LSP stars. We test whether the previously reported phase offset is a general property of LSPs. In addition, we compare this offset between different classes of long-period variables (LPVs).

\section{Observational Data}

\subsection{ASAS-SN photometry}
\label{sec:asassn}

The All-Sky Automated Survey for Supernovae (ASAS-SN) is a network of robotic telescopes that has been monitoring the sky for optical transients and variable sources since 2013 \citep{shappee2014}. The survey provides long-term photometric observations for sources across the entire sky and has produced catalogs of hundreds of thousands of variable stars. 
During its $V$-band phase, ASAS-SN monitored the sky down to $V\lesssim17$ mag with a typical cadence of approximately 2--3 days. Since 2018, the survey has primarily operated in the $g$ band, reaching $g\lesssim18.5$ mag with a cadence of $\lesssim24$~h \citep{christy2023}. All ASAS-SN light curves are publicly available to the astronomical community through an online interface\footnote{https://asas-sn.osu.edu/} \citep{kochanek2017}.

\subsection{{\it Gaia} radial velocities}
\label{sec:gaia}

The \textit{Gaia} mission provides high-precision astrometric, photometric, and spectroscopic observations for stars across the sky. As part of its Focused Product Release (FPR), \citet{gaiafpr2023} published RV curves for 9614 stars classified as LPVs, providing time-series spectroscopic data for studying the variability and orbital properties of pulsating red giants. This valuable dataset has already enabled further discoveries, including the identification of the longest-period classical Cepheid known in the Milky Way \citep{soszynski2024}.
RV curves available within the FPR were collected between July 2014 and May 2017, and contain between 12 and 90 points per star with a mean value of 24 observations. 

\subsection{Sample preparation}
\label{sec:sample}

\begin{figure*}[ht!]
\begin{center}
\begin{tabular}{rrr}
 \includegraphics[width=0.315\textwidth]{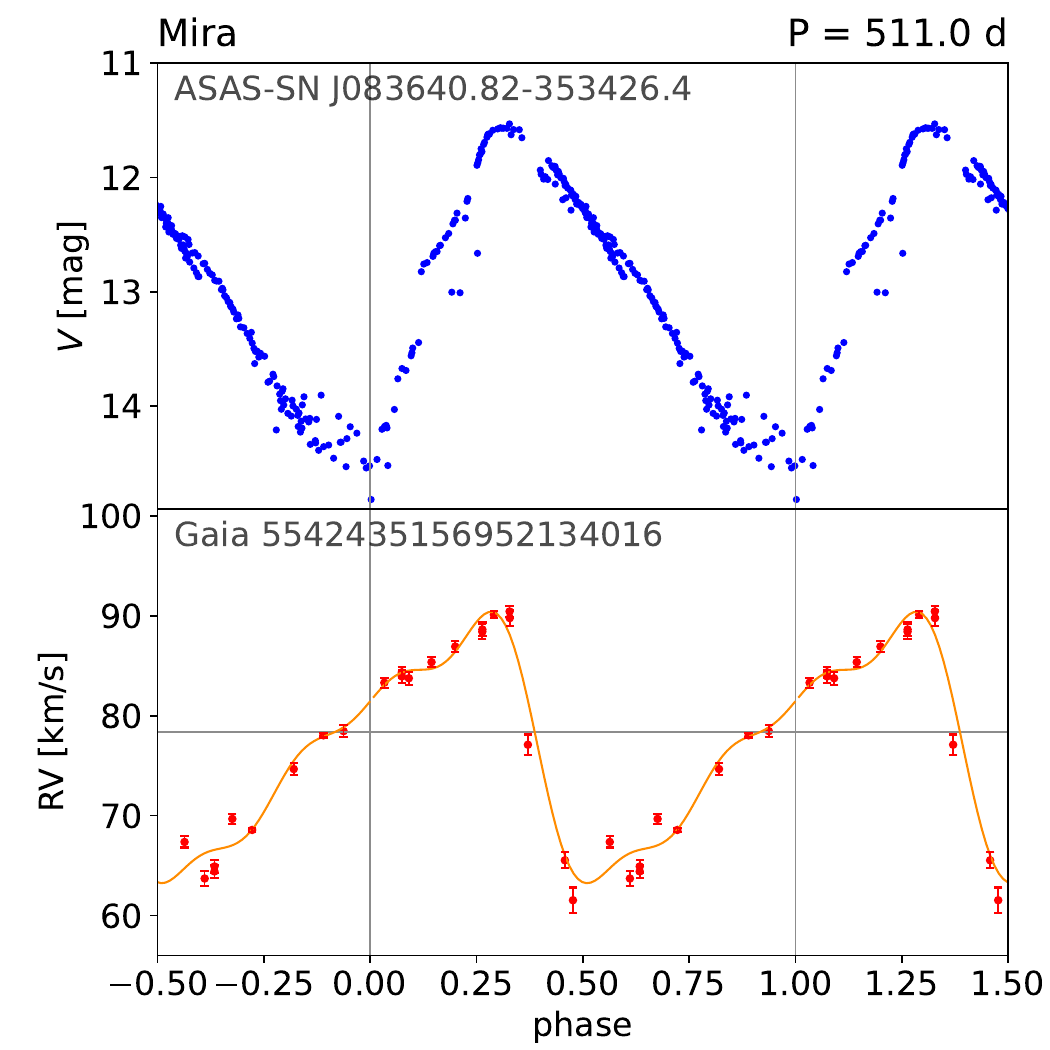} &
 \includegraphics[width=0.315\textwidth]{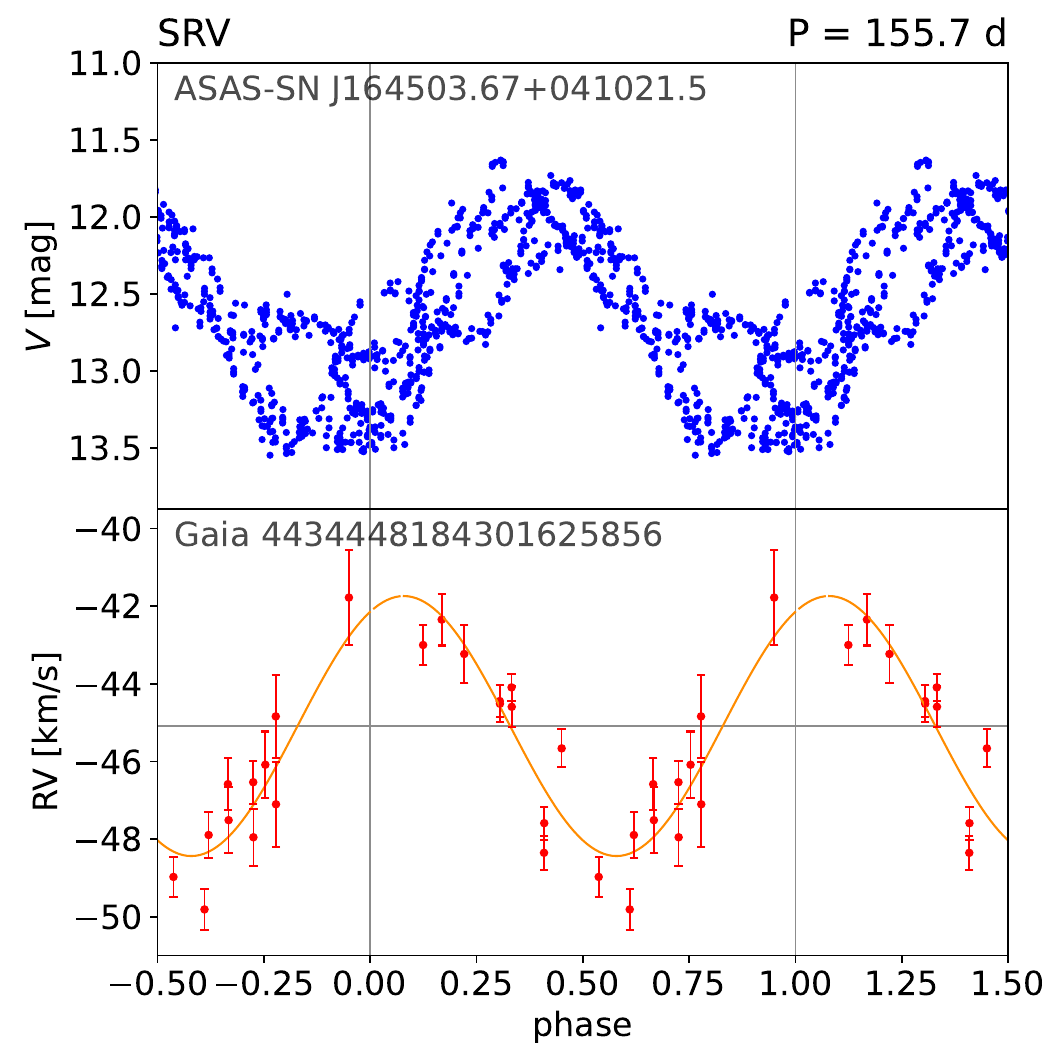}  &
 \includegraphics[width=0.315\textwidth]{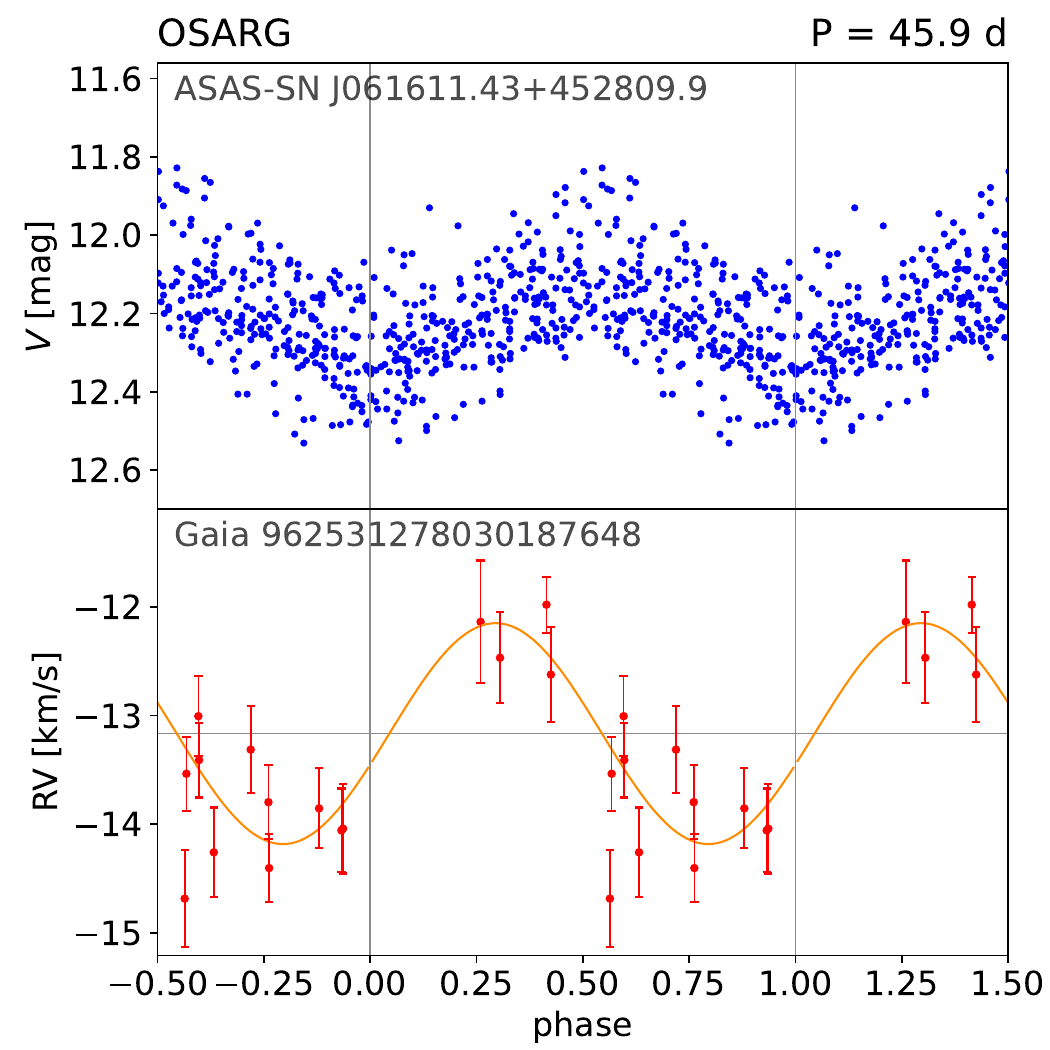} \\
 \includegraphics[width=0.31\textwidth]{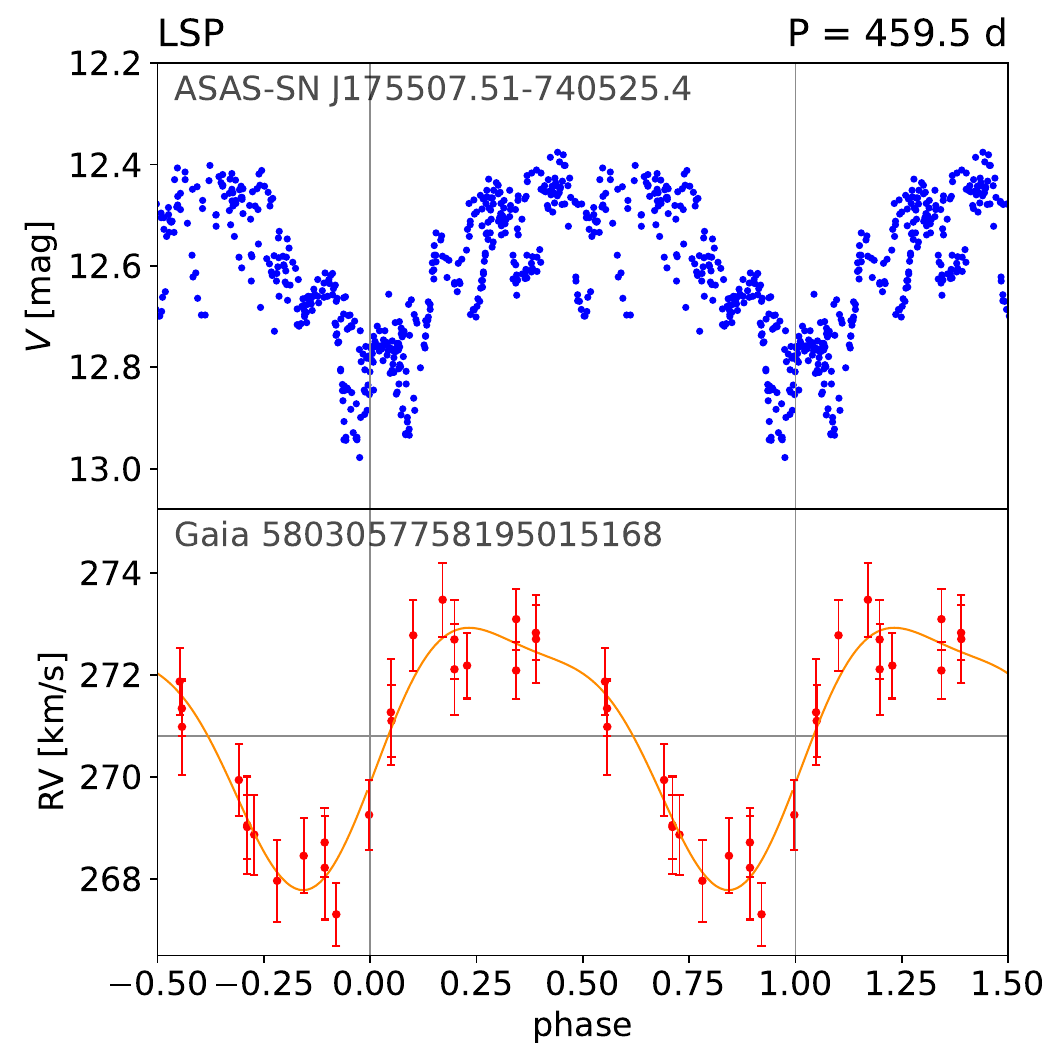} &
 \includegraphics[width=0.315\textwidth]{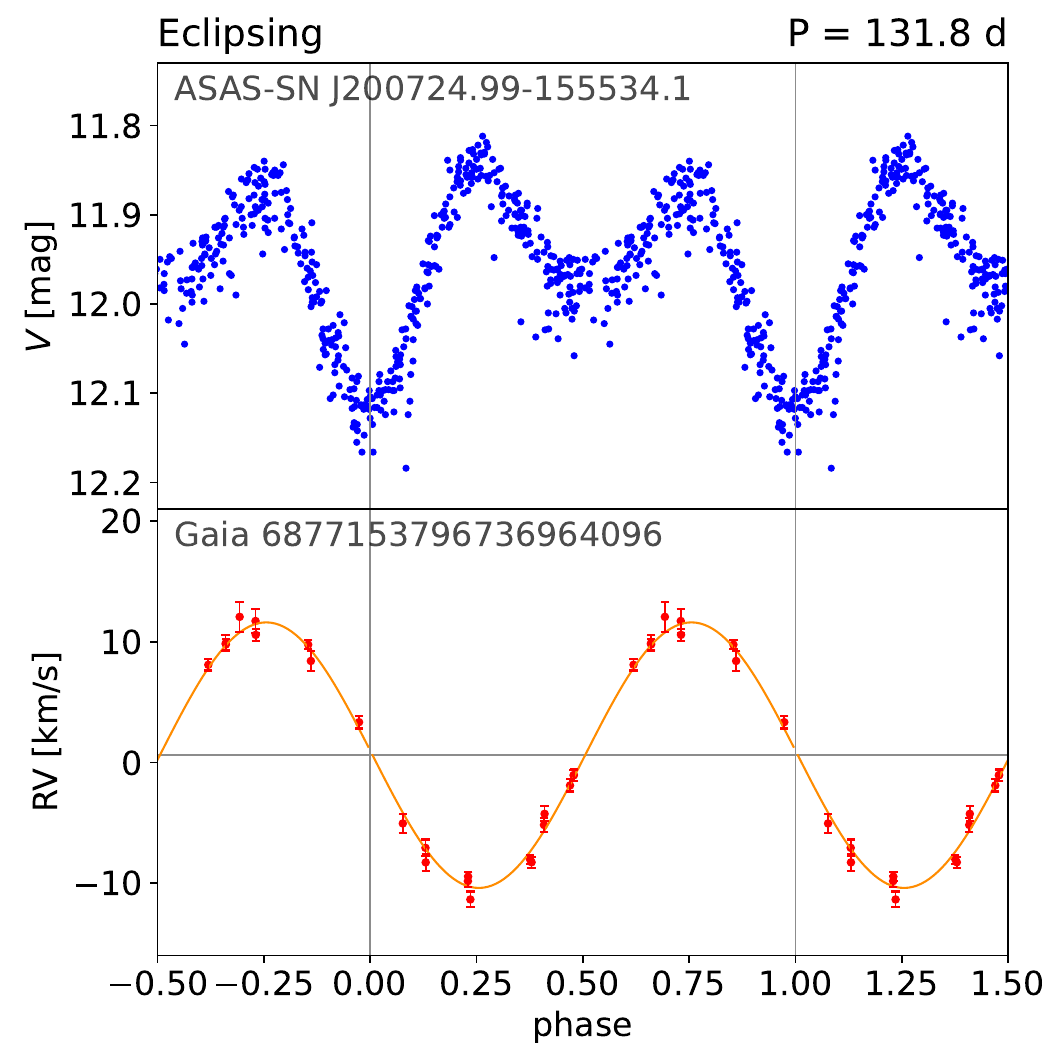} &
 \includegraphics[width=0.315\textwidth]{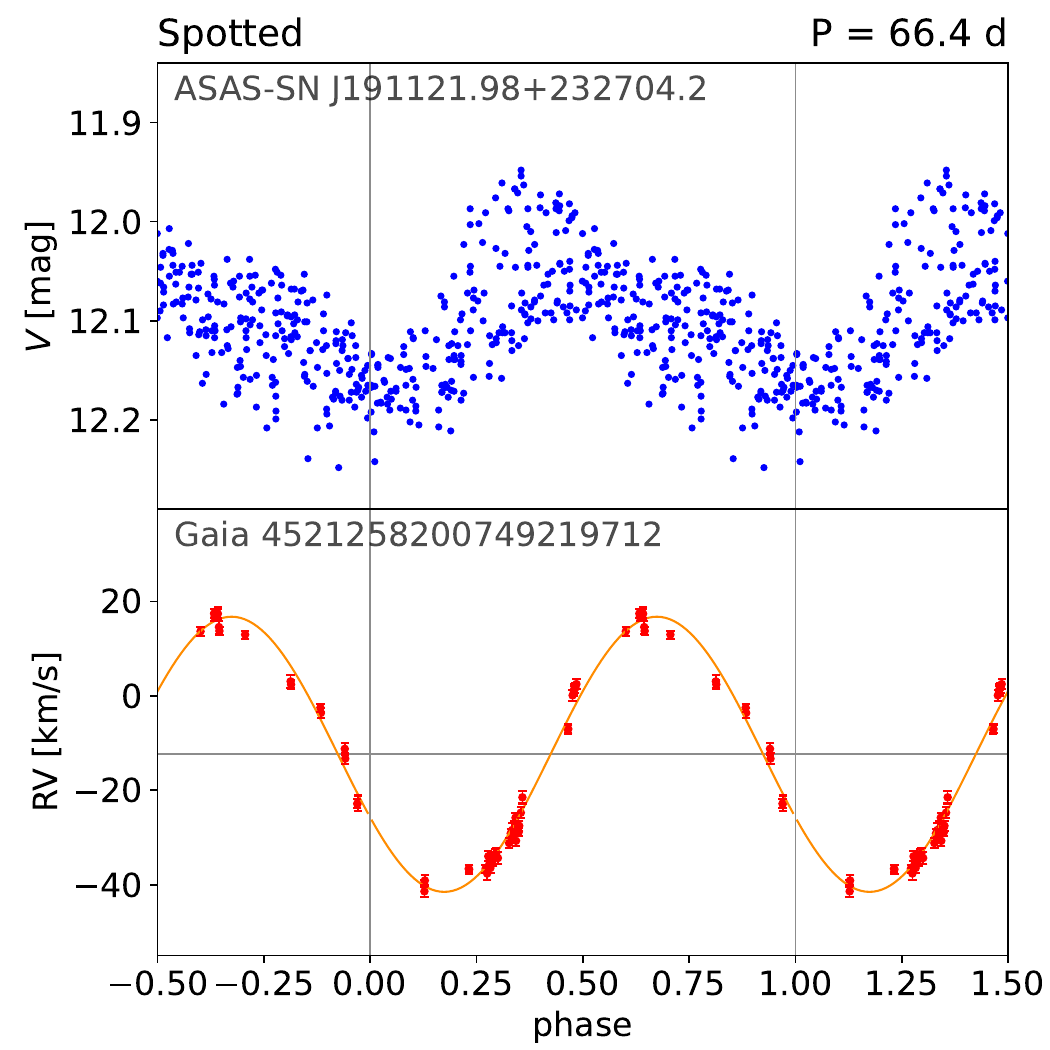} \\
\end{tabular}
\end{center}
\vspace{-10pt}
\caption{Each of the six panels shows a phase-folded ASAS-SN light curve (blue) and a {\it Gaia} RV curve (red) for six different variable classes, as indicated at the top of each panel, together with the pulsation period. Object IDs from both surveys are given within the panels. Grey vertical lines mark phases of the light curve minima (at 0.0 and 1.0), while the grey horizontal line marks the systemic velocity. The orange line shows the fit to the RV data, as described in the text.
\label{fig:variables}}
\end{figure*}

We cross-matched the {\it Gaia} FPR catalog of 9614 LPVs with the ASAS-SN catalog of variable stars and identified counterparts for 7960 objects. Their ASAS-SN {\it V}-band light curves were then examined for periodicity using the AOV and AOVTRANS algorithms implemented in the TATRY code \citep{schwarzenberg1996}. We visually inspected the light curves and classified them as Mira variables, semiregular variables (SRVs), OGLE Small Amplitude Red Giants (OSARGs), LSP variables, eclipsing and ellipsoidal binaries, spotted variables, or other types of variable stars (see \citealt{skowron2026} for a detailed description of these LPV classes). We included in our classification only those light curves that could be unambiguously assigned to one of these classes and for which the period of the brightness variations could be reliably determined. Variables with too few ASAS-SN measurements or an insufficiently long observational baseline were left unclassified. At this stage, we also excluded LPVs whose mean brightness showed significant irregular variations.

The final sample of 4130 stars consisted of 1887 SRVs, 983 LSP variables, 669 OSARGs, 312 eclipsing stars, 201 Miras, 44 spotted variables and 34 other variable stars which did not fit either of these categories. An example of each variable class is presented in Figure~\ref{fig:variables}.

During the visual inspection of the light and RV curves, we noticed that for many of the selected stars, the RV curves were either too noisy or contained too few measurements to reliably identify periodic variations. We therefore selected a subsample of Miras, SRVs, and LSP variables with well-sampled RV curves of sufficiently high quality, which we refer to as the {\em golden sample}. This sample consists of 40 Miras, 180 SRVs, and 100 LSP variables.

\section{Analysis}

\begin{figure*}[ht!]
\begin{center}
\includegraphics[width=0.85\textwidth]{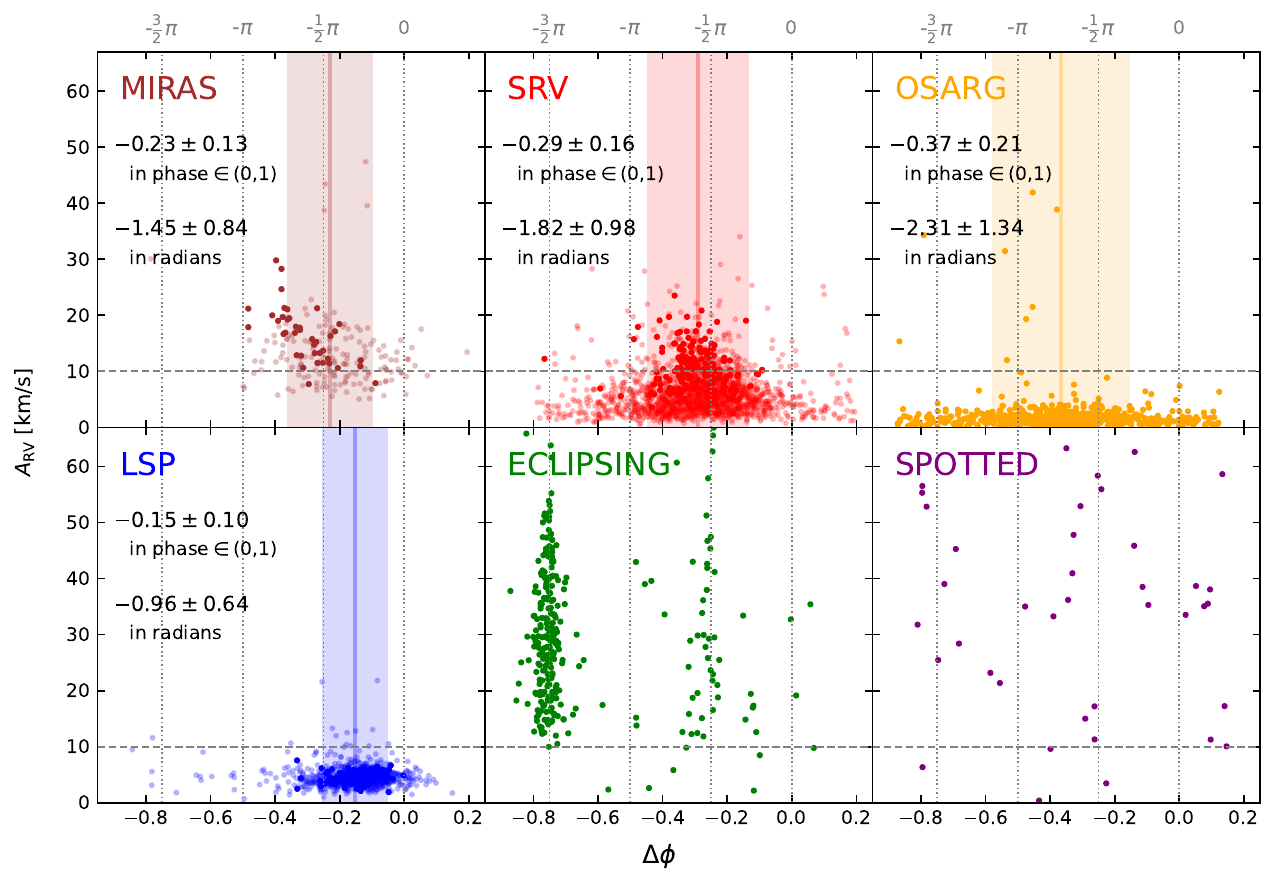}\\
\end{center}
\vspace{-15pt}
\caption{The amplitude of the RV curve ($A_{\rm RV}$) $vs.$ the difference between the minimum of the RV and light curve ($\Delta\phi$) for six LPV types. Darker points represent objects from the {\em golden sample}. The vertical lines show the mean value of the distribution of $\Delta\phi$, while the shaded regions mark the $\pm 1\sigma$ values around the mean, for Miras, SRVs, OSARGs and LSPs. The grey vertical dotted lines mark phases of $-\frac{3}{2}\pi$, $-\pi$, $-\frac{1}{2}\pi$, and $0$ rad, as marked at the top of the plot.
\label{fig:ampl_phase}}
\end{figure*}

The variability period determined from the ASAS-SN light curve was used to phase both the photometric and {\it Gaia} RV observations. For each star, we determined the epoch of minimum brightness, $T_\mathrm{min}^V$, and the epoch of minimum radial velocity, $T_\mathrm{min}^\mathrm{RV}$, by fitting a third-order Fourier series to the light curve and a first-order Fourier series (a sinusoid) to the RV curve of OSARGs, eclipsing and spotted variables, and a second-order Fourier series to those of Miras, SRVs and LSPs. The choice of the fitting order was based on the typical shape of the RV curves within each group of variables. We then defined the phase offset as \(\Delta\phi = (T_\mathrm{min}^\mathrm{RV}-T_\mathrm{min}^V)/P\), where $P$ is the period. The
phase offset can be expressed either in cycles or in radians as $2\pi\Delta\phi$. This convention allows a direct comparison with the RV–photometric phase relation discussed by \cite{goldberg2024}.

In Figure~\ref{fig:ampl_phase} we plot the peak-to-peak amplitudes of the RV curves, ${A_{\rm RV}}$, as a function of the phase offsets $\Delta\phi$, for the six variable star classes. The mean phase offsets and their $\pm 1\sigma$ values are shown in the figure and summarized in Table~\ref{tab:offsets} for the four classes of interest: Miras, SRVs, OSARGs, and LSPs.

\begin{deluxetable}{lrcc}[hbt]
\tablewidth{0pt}
\tablecaption{Mean phase offsets between the phase of the minimum of the RV and light curve. The uncertainties represent the standard deviations of the phase offset distributions.\label{tab:offsets}}
\tablehead{
\colhead{Variable} & \colhead{\# stars} & \multicolumn{2}{c}{$\Delta\phi$} \\
\vspace{-3pt}
 &  & \colhead{in phase$\in$(0,1)} &  \colhead{in radians} 
}
\startdata
Mira &  201 & $-0.23\pm0.13$ & $-1.45\pm0.84$ \\
SRV   & 1887 & $-0.29\pm0.16$ & $-1.82\pm0.98$ \\
OSARG &  669 & $-0.37\pm0.21$ & $-2.31\pm1.34$ \\
LSP   &  983 & $-0.15\pm0.10$ & $-0.96\pm0.64$ \\
\enddata
\tablecomments{Third-order Fourier fit to the light curves and first (OSARGs) or second (Miras, SRVs and LSPs) order Fourier fit to the RV curves. Eclipsing and spotted variables are omitted, as their mean phase offsets are not considered informative in the context of this analysis.}
\end{deluxetable}

Miras are AGB stars pulsating in the radial fundamental mode with high-amplitude light and RV curves, making the estimation of $\Delta\phi$ subject to relatively low uncertainties. SRVs usually pulsate in at least two radial modes, most often the fundamental and first overtone, resulting in more irregular light and RV curves. This also leads to lower and noisier RV amplitudes, making it more difficult to determine $T_\mathrm{min}^\mathrm{RV}$ and contributing to the broader distribution of $\Delta\phi$.

The distribution of OSARGs is even broader. These stars are less luminous than Miras or SRVs and pulsate in a mixture of radial and non-radial modes. As a result, their RV curves are typically the noisiest, making the determination of $T_\mathrm{min}^\mathrm{RV}$ highly uncertain (the OSARG RV curve shown in Figure~\ref{fig:variables} is an example of one of the few higher-quality RV curves). Their RV amplitudes are also the lowest, with a median of $1.5$~km~s$^{-1}$.  

\begin{figure*}[ht]
\begin{center}
\includegraphics[width=0.95\textwidth]{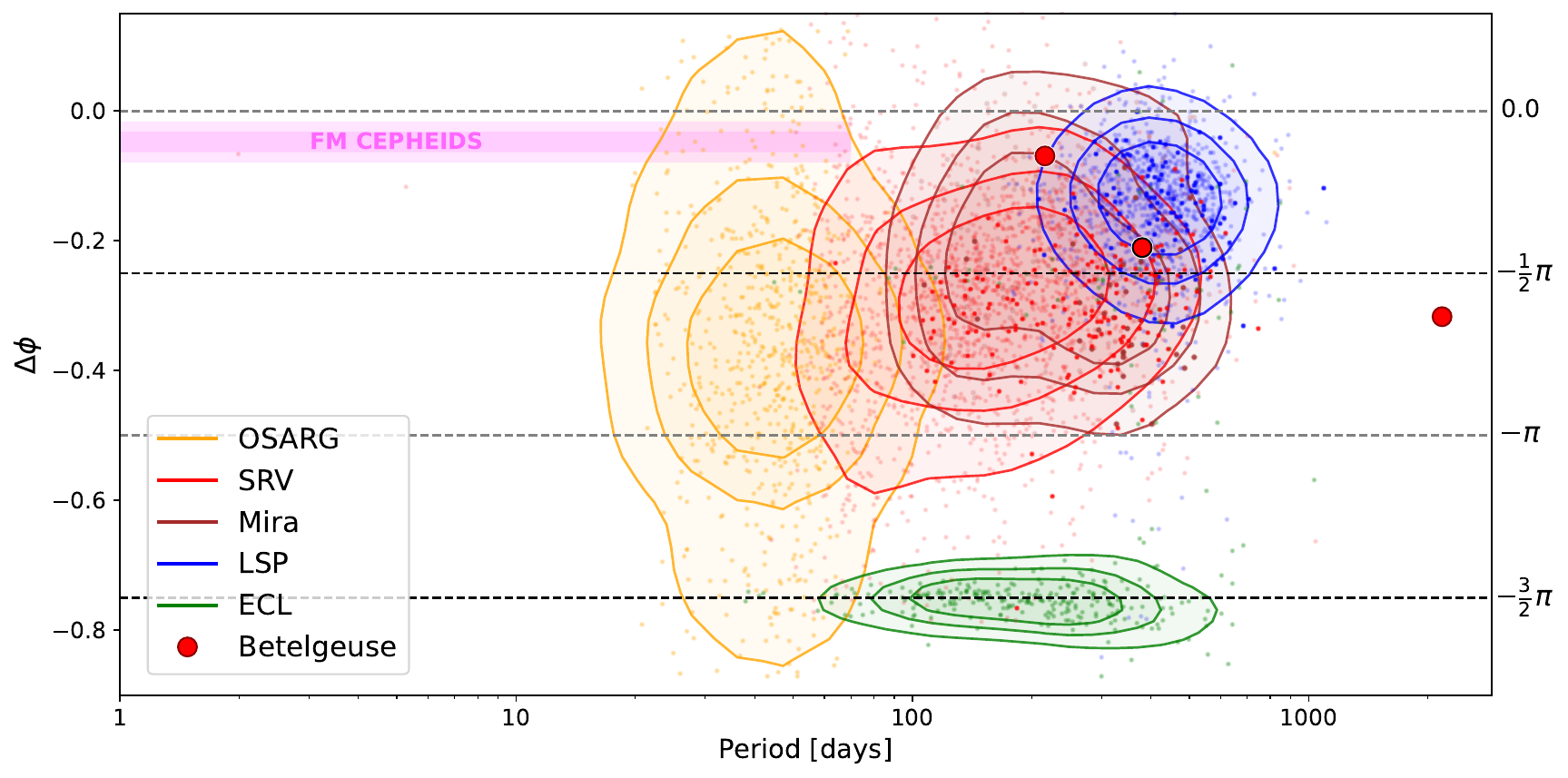}
\end{center}
\vspace{-5pt}
\caption{
Phase offset between the RV and light curves as a function of variability period for different classes of variable stars, in a form analogous to Figure 3 of \citet{goldberg2024}. Individual sources are shown as semi-transparent points, while darker points represent objects from the {\em golden sample}. Contours indicate normalized density levels at 15\%, 35\%, and 60\% of the peak density for each class. The pink shaded region marks the approximate location of fundamental-mode classical Cepheids. The large red dots indicate the three variability periods and corresponding phase offsets of Betelgeuse reported by \citet{goldberg2024}. The horizontal dashed lines mark phase offsets of $0$, $-\frac{1}{2}\pi$, $-\pi$, and $-\frac{3}{2}\pi$ rad.
\label{fig:phase_per}}
\end{figure*}

LSP stars, on the other hand, occupy a fairly compact region in the $\Delta\phi$\,--\,$A_{\rm RV}$ plane, with a mean $\Delta\phi$ value similar to that of Miras, but with much lower RV amplitudes (the median $A_{\rm RV}$ is $11.1$~km~s$^{-1}$ for Miras and only $3.9$~km~s$^{-1}$ for LSP stars). The similarity of their phase offsets might suggest a common origin of both types of variability. However, radial pulsations, which drive Mira variability, have already been excluded as an explanation for LSPs \citep{goldberg2024}. 
Thus, the phase offset alone cannot distinguish LSP variability from radial pulsations, particularly those in the fundamental mode.


In the case of eclipsing variables, the phase offset is, as expected for circular orbits, at $\Delta \phi = -0.75 = -\frac{3}{2}\pi$ (or at $-0.25 = -\frac{1}{2}\pi$ depending on the temperatures of stars), meaning that at minimum light, the companion is between the observer and the host star (or behind it), and the relative RV is equal to zero. The RV amplitudes span a wide range of values, typically between $10$ and $60$~km~s$^{-1}$, indicating a wide range of companion masses.
In the case of spotted variables, there is no particular preference for $\Delta \phi$, and both $\Delta \phi$ and $A_{\rm RV}$ show broad, apparently random distributions. Photometric studies of spotted variables have revealed a wide diversity of the properties of light curves, reflecting the complex nature of stellar magnetic activity \citep{iwanek2019, iwanek2024}. At the same time, the phase relation between photometric and RV variations produced by spots depends on the spot geometry and stellar inclination. Therefore, the absence of a well-defined phase relation between photometric and RV variations is expected.

Figure~\ref{fig:phase_per} shows the phase offsets as a function of variability period, in a form analogous to Fig.~3 of \citet{goldberg2024}. These authors found that LSP stars are concentrated around $\Delta\phi\sim-\pi/2$.
Our much larger sample confirms that LSP stars occupy a relatively narrow range of phase offsets, but with a distribution centered closer to $\sim-\pi/3$ than to $\sim-\pi/2$. Part of this difference can be attributed to our use of a second-order Fourier series to fit the non-sinusoidal RV curves, rather than a simple sinusoid. Using a first-order Fourier series to fit the RV curves, we obtain a mean phase offset of $\Delta\phi = -0.36 \pi$, somewhat closer to the value reported by \cite{goldberg2024}.
Figure~\ref{fig:phase_per} also demonstrates that this phase relation is not unique to LSPs. In particular, Miras show significant overlap with LSP stars in $\Delta\phi$, despite occupying a different range of periods. SRVs and OSARGs extend toward more negative phase offsets and shorter periods. 

Figure~\ref{fig:phase_per} also includes the three periods and corresponding phase offsets reported for Betelgeuse by \citet{goldberg2024}, associated with its first-overtone, fundamental-mode, and LSP variability. These measurements follow a trend similar to that observed for the other LPVs. The first overtone phase offset is consistent with those of shorter period pulsators, while the LSP phase offset falls close to the distribution of LSP stars. The fundamental mode lies away from the region occupied by fundamental mode classical Cepheids, shown schematically following \citet{goldberg2024}, but its phase offset is consistent with the distribution of Miras. Thus, the apparent displacement of Betelgeuse's fundamental-mode phase offset from the classical Cepheid sequence is not unusual when compared with other luminous red giants pulsating in the fundamental mode.

\begin{figure}[t!]
\includegraphics[width=0.47\textwidth]{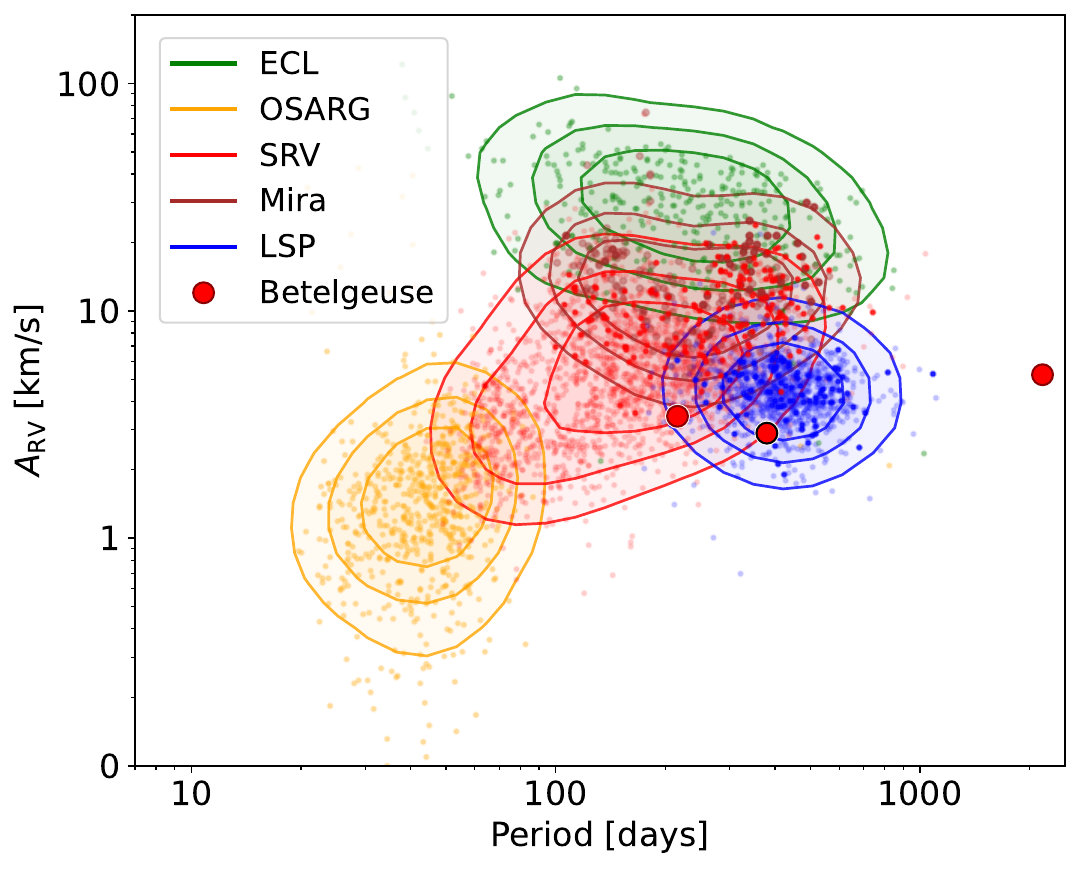}
\vspace{-15pt}
\caption{Period–RV amplitude diagram for OSARGs, SRVs, Mira variables, LSP stars, and eclipsing binaries. Individual sources are shown as semi-transparent points, while darker points represent objects from the {\em golden sample}. Contours trace normalized density levels at 15\%, 35\%, and 60\% of the peak density of each class. The large red dots indicate the periods and corresponding radial-velocity amplitudes of Betelgeuse reported by \cite{goldberg2024}.
\label{fig:ampl_per}}
\end{figure}

Although the phase offset alone does not clearly separate LSP stars from pulsating red giants, the difference becomes clearer when the RV amplitude is considered together with the variability period (Figure~\ref{fig:ampl_per}). LSP stars form a relatively compact group at long periods and low RV amplitudes, typically of only a few~km~s$^{-1}$. In contrast, OSARGs, SRVs, and Miras appear to follow a general trend of increasing RV amplitude with period, while LSP stars remain at relatively low RV amplitudes despite their much longer periods. This provides yet another indication that LSPs do not represent a simple extension of the ordinary pulsational variability of red giants toward longer periods, although  other types of non-radial pulsations cannot be entirely excluded.
At the same time, RV amplitudes of a few~km~s$^{-1}$ are consistent with the orbital motion expected for a brown dwarf or very low-mass stellar companions on orbits with radii of the order of AU \citep{nicholls2009,decin2025}.

\begin{figure*}[t!]
\begin{center}
\includegraphics[width=0.32\textwidth]{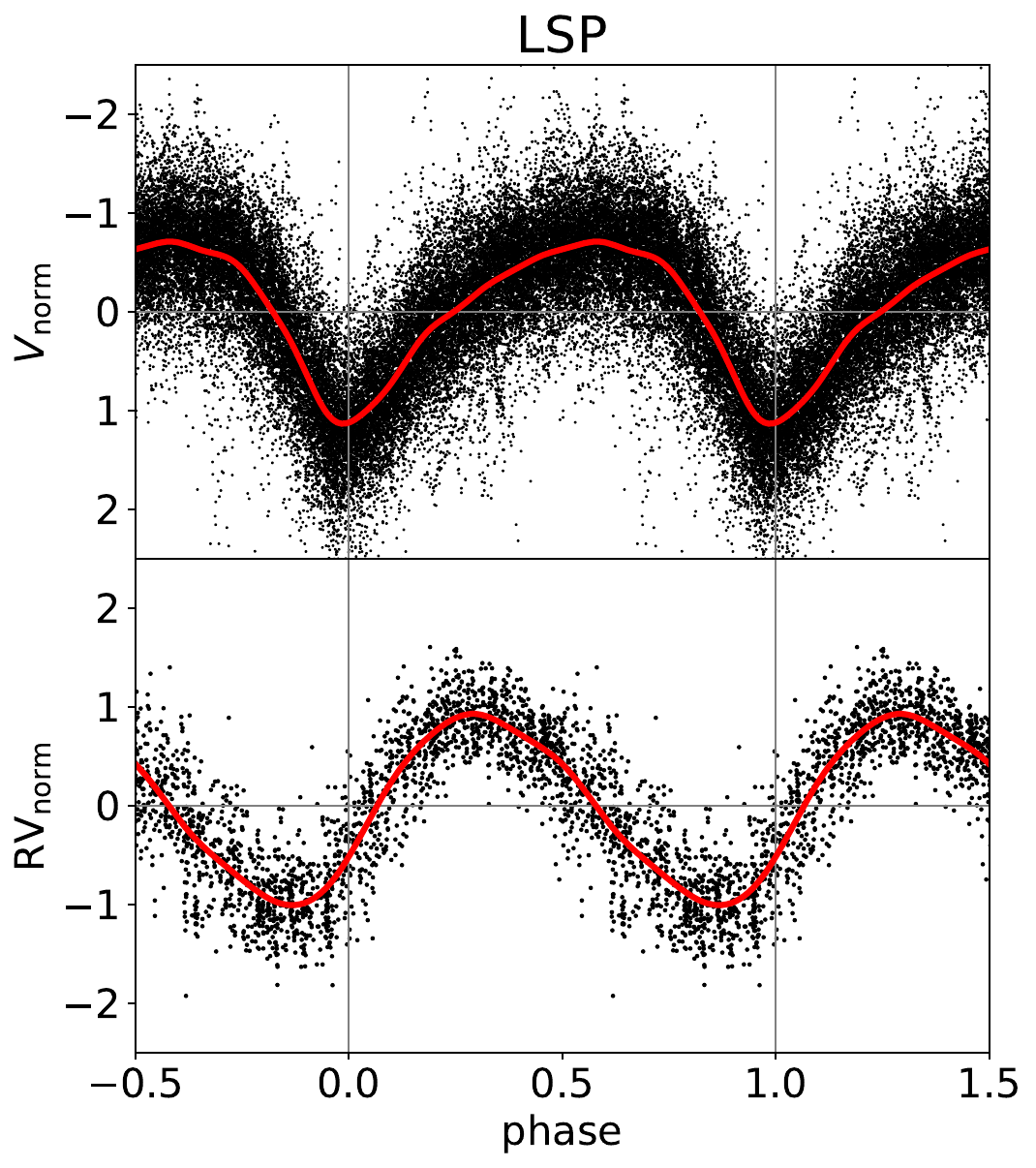}
\includegraphics[width=0.32\textwidth]{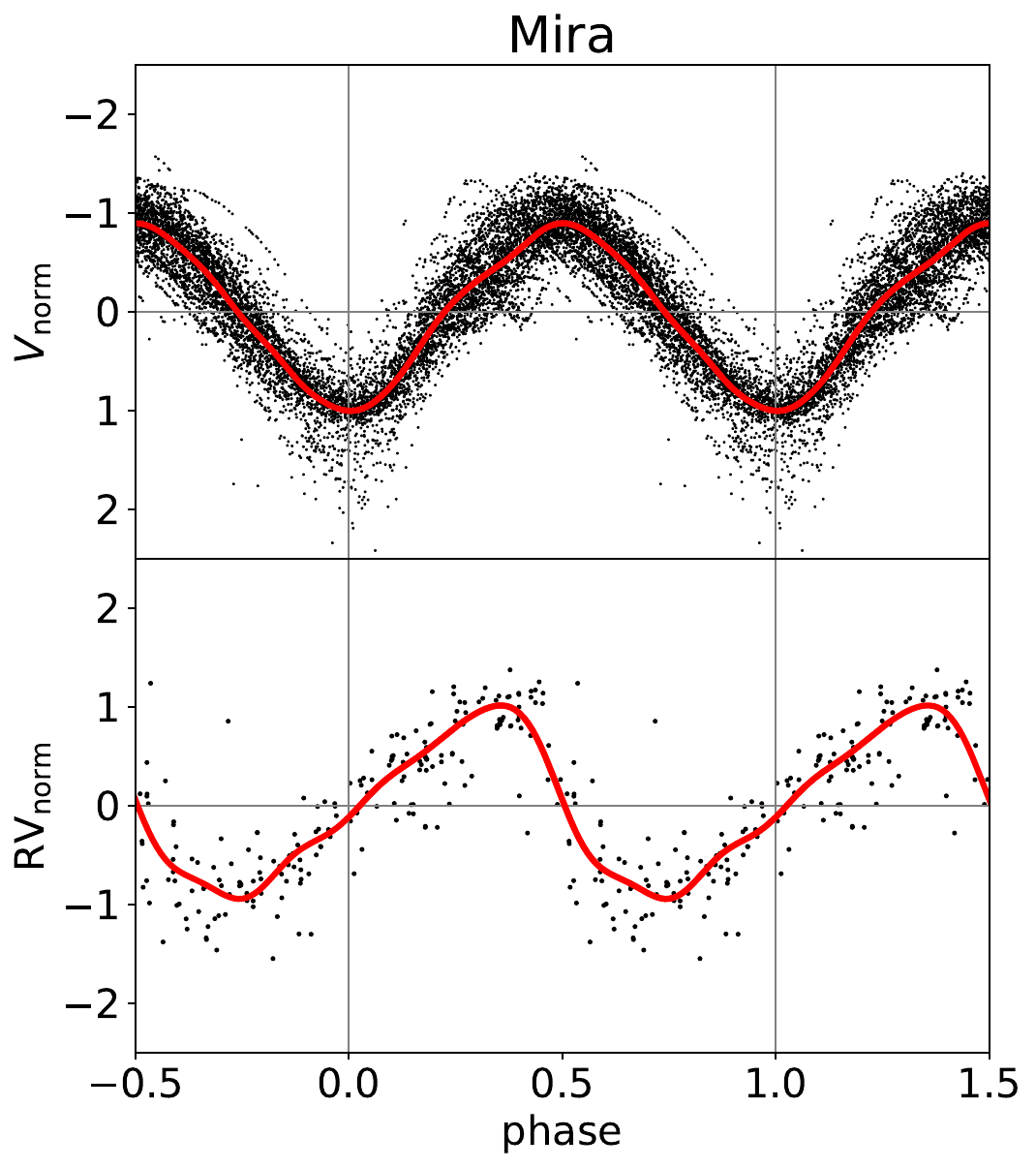}
\includegraphics[width=0.32\textwidth]{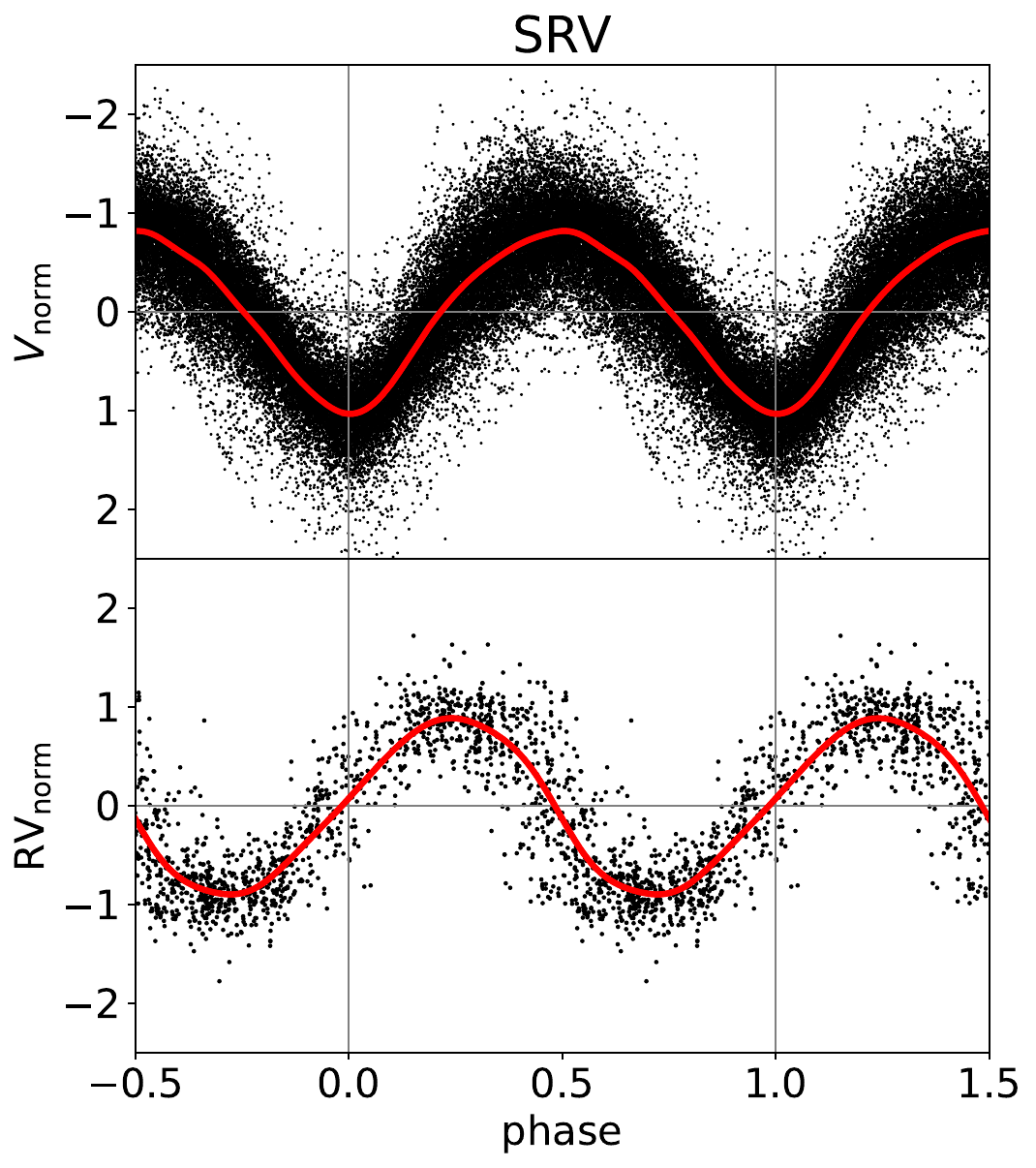}
\end{center}
\vspace{-5pt}
\caption{Normalized $V$-band light (top) and RV (bottom) curves of the {\em golden sample} of 100 LSP variables (left), 40 Miras (middle), and 180 SRVs (right) stacked on top of each other. Red lines show the averaged data. 
\label{fig:averages}}
\end{figure*}

Figure~\ref{fig:averages} shows the normalized light and RV curves of stars from the {\em golden sample} stacked together with their mean curves overplotted. This provides another illustration of the remarkable homogeneity of the light and RV curve shapes and the coherent phase relation between the photometric and RV variations among LSP stars. Despite the large scatter in the individual measurements, the averaged curves reveal a very clear and systematic phase relation, demonstrating that it is a general property of the LSP population and therefore providing an important constraint on the origin of LSPs.
As pointed out by \cite{soszynski2021} and \cite{goldberg2024}, this phase relation is not consistent with a simple binary model in which the obscuring material is located at the position of the companion. Instead, the material producing the photometric minimum has to be shifted in orbital phase with respect to the companion. This may naturally occur if the companion produces a dusty wake in the wind of the red giant. The hydrodynamical simulations of \citet{decin2025} show that the maximum optical depth can be shifted by about $90^\circ$--$225^\circ$ from the companion, producing phase offsets consistent with those observed for LSP stars. 
Importantly, such a model naturally allows a range of phase offsets, rather than requiring a single fixed phase relation. However, the relatively narrow observed distribution provides a useful additional constraint on the range of wake geometries, orbital parameters, and viewing angles capable of producing detectable LSP variability.
Moreover, the very different shapes of the mean light and RV curves compared with those of pulsating red giants (Miras and SRVs) provide further evidence against pulsations as the origin of LSPs and impose an additional observational constraint on models of the LSP phenomenon.

Taken together, these results provide strong constraints on the origin of LSPs. Although the phase offset alone cannot distinguish LSP stars from pulsating red giants, the combination of long periods and low RV amplitudes clearly separates them from other pulsating LPVs. In the binary interpretation, the small RV amplitudes are consistent with low-mass companions, while the systematic phase offset requires the obscuring material to be shifted in orbital phase relative to the companion. The coherent phase relation observed across our large sample is consistent with this interpretation, while its relatively narrow distribution provides an additional constraint on the geometry of the obscuring material.
A dusty wake produced by the interaction of such a companion with the wind of the red giant can account for both properties, as shown by the hydrodynamical simulations of \cite{decin2025}. The distinct shapes of the mean light and RV curves provide further constraints that models of LSP variability must reproduce.

\section{Summary and Conclusions}

We investigated the phase relation between the light and RV curves of 983 LSP stars using ASAS-SN photometry and {\it Gaia} FPR RV data. For comparison, we performed the same analysis for Miras, SRVs, OSARGs, eclipsing binaries, and spotted variables. We confirm the systematic phase offset previously reported for LSP stars, but with a much larger sample. The distribution of phase shifts in LSP variables is relatively narrow, with a mean value of $\Delta\phi=-0.15\pm0.10$ in phase ($-0.96\pm0.64$ rad), demonstrating that this phase relation is a general property of LSP stars.

Interestingly, the phase offsets of LSP stars are very similar to those of Miras, which pulsate in the radial fundamental mode. This shows that the phase offset alone cannot distinguish between LSP and pulsational variability. However, LSP stars behave differently from the pulsating red giants in the period--RV amplitude plane. OSARGs, SRVs, and Miras appear to follow a general trend of increasing RV amplitude with period, while LSP stars remain at relatively low amplitudes of only a few~km~s$^{-1}$ despite their much longer periods. Thus, the combination of period and RV amplitude provides a clearer distinction between LSP stars and pulsating LPVs than the phase offset alone. Together with the fact that LSP periods are several times longer than the fundamental mode periods, this argues against interpreting LSPs as an additional 'ordinary' pulsation mode. 

The stacked light and RV curves of the golden sample further show how coherent the phase relation is among LSP stars. The persistence of this relation confirms that it is a general property of LSP variability.
In a binary interpretation, the observed phase relation cannot be reproduced if the obscuring material is located at the position of the companion \citep{soszynski2021,goldberg2024}. Instead, the obscuring material has to be shifted in orbital phase with respect to the companion. Such a shift can be produced by a dusty wake formed through the interaction of a low-mass companion with the wind of the red giant, as shown by the hydrodynamical simulations of \citet{decin2025}. The narrow distribution of phase offsets, together with the characteristic shapes of the mean light and RV curves, provide additional observational constraints for such models.

\begin{acknowledgments}
DMS, PI, and PKS acknowledge support from ERC grant LSP-MIST (Grant Agreement No. 101040160). Funded by the European Union. Views and opinions expressed are however those of the authors only and do not necessarily reflect those of the European Union or the European Research Council Executive Agency. Neither the European Union nor the granting authority can be held responsible for them. This work has been supported by the National Science Centre, Poland, through grant 2022/45/B/ST9/00243 to IS.

This work has made use of data from the European Space Agency (ESA) mission {\it Gaia} (\url{https://www.cosmos.esa.int/gaia}), processed by the {\it Gaia} Data Processing and Analysis Consortium (DPAC, \url{https://www.cosmos.esa.int/web/gaia/dpac/consortium}). Funding for the DPAC has been provided by national institutions, in particular the institutions participating in the {\it Gaia} Multilateral Agreement.
\end{acknowledgments}

\facilities{ASAS-SN, {\it Gaia}}

\software{NumPy \citep{harris2020array}, Matplotlib \citep{hunter2007}}

\bibliography{paper}
\bibliographystyle{aasjournalv7}

\end{document}